\documentclass{ws-procs9x6}

\def\tt{t\bar t}
\def\pp{p\bar p}

\def\roots{\sqrt{s}}
\def\gammax{\Gamma_X}

\def\met{\mbox{${\hbox{$E$\kern-0.6em\lower-.1ex\hbox{/}}}_T$}}
\def\metcal{\mbox{${\hbox{$E$\kern-0.6em\lower-.1ex\hbox{/}}}_T$}^{cal}}

\newcommand\degree{^\circ}                      %degree sign

\def\gevcc{{\rm GeV/c^2}}
\def\chisq{\chi^2}

\newcommand\eqref[1]{Eq.~(\ref{#1})}

\newcommand\figref[1]{Fig.~\ref{#1}}
\newcommand\tabref[1]{Table~\ref{#1}}

\def\chisq{\chi^2}

\begin{document}
\title{Search for Narrow Width $\tt$ Resonances in $\pp$ collisions at $\roots$ = 1.8 TeV.}
\author{Supriya Jain*, Naba Mondal, and Dhiman Chakraborty}
\address{*E-mail: sjain@tifr.res.in}
\author{(for the D$\O$ Collaboration)}


\maketitle

\abstracts{
We present a preliminary result on a search for narrow width resonances that decay into $\tt$ pairs using 130 $\rm{pb^{-1}}$ of lepton plus jets data in $\pp$ collisions at $\roots$ = 1.8 TeV. No significant deviation from Standard Model prediction is observed. 95$\%$ C.L. upper limits on the production cross section of the narrow width resonance times its branching fraction to $\tt$ are presented for different resonance masses $M_X$. We also exclude the existence of a leptophobic topcolor particle, $X$, with $M_X$ $<$ 560 $\gevcc$ for a width $\gammax$ = 0.012$M_X$.
}

%%%%%%%%%%%%%%%%%%%%%%%%%%%%%%%%%%%%%%%%%%%%%%%%%%%%%%%%%%%%%%%%%%%%%%%
%  1st page...........
%%%%%%%%%%%%%%%%%%%%%%%%%%%%%%%%%%%%%%%%%%%%%%%%%%%%%%%%%%%%%%%%%%%%%%

Particles with narrow width that decay to $\tt$ pairs are predicted by several non Standard Model theories~\cite{topcolor,hillparkeharris}. For instance, in one of the scenarios of the topcolor-assisted technicolor model in Ref. [2], a heavy $Z'$ is predicted, that couples preferentially to the third quark generation. \\

 At present, direct searches for these heavy particles or resonances are 
 possible only at the Tevatron, the 1.8 TeV $\pp$ collider located at the Fermi National Accelerator Laboratory. Experiments seek an excess, beyond that predicted by the Standard Model (SM), in the distribution of the invariant mass of the $\tt$ decay products. Previous searches~\cite{cdf} from the Tevatron have limited a leptophobic $Z'$ to a mass higher than 480 $\gevcc$. In this paper we present a preliminary result based on a direct search for $\tt$ narrow width, heavy resonances \hspace{0.5mm}in \hspace{0.5mm}the \hspace{0.5mm}inclusive \hspace{0.5mm}decay \hspace{0.5mm}modes \hspace{0.5mm}$\tt~\rightarrow~\ell~\nu$ + 4 (or more) jets, where $\ell~=$ $e$ or $\mu$, using 130 $\rm{pb^{-1}}$ of data recorded from 1992 to 1996
by the D$\O$ experiment at the Tevatron. \\

We consider two orthogonal classes of events for this analysis, whose selection is based on: a) a purely topological selection of lepton+jets events which we denote as $e+jets$ and $\mu+jets$, and b) a selection based primarily on the presence of a non-isolated, soft muon ($\mu$ tag) from $b$ and $c$ quark semileptonic decays, with additional selections on the topology of the event. These events are denoted as $e+jets/\mu$ and $\mu+jets/\mu$. The principal sources of background are due to SM $\tt$ production, production of {\it{W}}($\rightarrow$ $l$$\nu$) + $\ge$ 4 jets, and production of multijets $(N_{j} \sim 5)$, in which one of the jets is misidentified as a lepton, and instrumental effects simulate sufficient $\met$ satisfying the neutrino requirement. The selection criteria used to reduce the contribution from non-$\tt$ sources are summarized in \tabref{tb:prcuts}.   \\
 
%%%%%%%%%%%%%%%%%%%%%%%%%%%%%%%%%%%%%%%%%%%%%%
% SELECTIONS
%%%%%%%%%%%%%%%%%%%%%%%%%%%%%%%%%%%%%%%%%%%%%%
\begin{table}[ph]
\tbl{Summary of event selections.\vspace*{1pt}} 
{\footnotesize
\begin{tabular}{|c|c|c|c|c|}
\hline
{} &{} &{} &{} &{}\\[-1.5ex] 
& $e$+jets & $\mu$+jets &  $e$+jets/$\mu$ &  $\mu$+jets/$\mu$\\ [1ex]
\hline
{} &{} &{} &{} &{}\\[-1.5ex]
Lepton& $E_T>$20 GeV  & $p_T>$20 GeV/c  &  $E_T>$20 GeV &$p_T>$20 GeV/c  \\[1ex]
& $|\eta|<$2 & $|\eta|<$1.7 & $|\eta|<$2 &$|\eta|<$1.7 \\[1ex]\hline
$\met$&$\met >$20 GeV & $\met >$20 GeV & $\met >$20 GeV & \hspace*{-.5cm} 
$\met >$20 GeV \\[1ex]\hline
Jets& $\ge$ 4 jets, $|\eta|<$2 & $\ge$ 4 jets, $|\eta|<$2 & $\ge$ 4 jets, $|\eta|<$2 &  $\ge$ 4 jets, $|\eta|<$2 \\[1ex]
& $E_T>$15 GeV & $E_T>$15 GeV & $E_T>$15 GeV &$E_T>$15 GeV \\[1ex]\hline
$\mu$ tag& No & No & Yes & Yes \\[1ex]\hline

Other& $E_T^W>$60 GeV & $E_T^W>$60 GeV &$\met >$35 GeV, if&$\Delta\phi (\met,\mu) 
< 170 \degree$, if \\[1ex]
& $|\eta^W|<$2  & $|\eta^W|<$2& $\Delta\phi (\met,\mu) < 25\degree$& $\frac{|
\Delta\phi(\met,\mu)-90\degree|}{90\degree}$$<\frac{\met}{45GeV}$\\[1ex]\hline
Events & & & & \\[1ex]
selected & 42 & 41 & 4 & 3 \\[1ex] \hline
\end{tabular}\label{tb:prcuts} }
\vspace*{-13pt}
\end{table}

We consider the resonance signal ($X$ $\rightarrow$ $\tt$) at nine different masses $M_X$ between 400-1000 $\gevcc$, with a natural width $\gammax$ = 0.012$M_X$. We perform a three constraint kinematic fit to the $\tt$ $\rightarrow$ $l$ + jets, decay hypothesis~\cite{myprl}, and apply a cut of $\chisq$$<$10 to further reduce non-$\tt$ background, whereupon 41 events are left in the data sample of which 4 are $\mu$-tagged. \\
 
%%%%%%%%%%%%%%%%%%%%%%%%%%%%%%%%%%%%%%%%%%%%%%
% FITTING OF DATA TO A THREE SOURCE MODEL
%%%%%%%%%%%%%%%%%%%%%%%%%%%%%%%%%%%%%%%%%%%%%%

We then use Bayesian statistics~\cite{bayes} to fit the data $m_{\tt}$ distribution to a three-source model comprising signal ($X$ $\rightarrow$ $\tt$) and the SM backgrounds~\cite{myprl}. No significant deviation is seen in the data $m_{\tt}$ distribution from SM expectations for any of the resonance masses considered. \\

%%%%%%%%%%%%%%%%%%%%%%%%%%%%%%%%%%%%%%%%%%%%%%
% CONCLUSIONS
%%%%%%%%%%%%%%%%%%%%%%%%%%%%%%%%%%%%%%%%%%%%%%
 
\begin{figure}[ht]
\centerline{\epsfxsize=2.5in\epsfbox{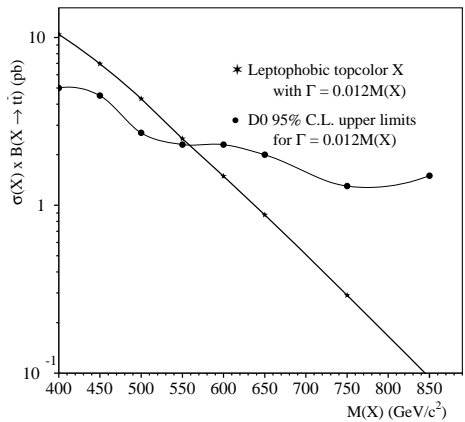}}   
\caption{The D$\O$ Run I 95$\%$ confidence level upper limits on $\sigma_X B$ as a function of resonance mass $M_X$. Included for reference are the predicted topcolor assisted technicolor cross sections for a width $\Gamma_X$ = 1.2$\%$ $M_X$. \label{mxbound}}
\end{figure}

To conclude, after investigating 130 $\rm{pb^{-1}}$ of data, we find no statistically significant evidence for $\tt$ resonance, and so establish upper limits on $\sigma_X B$($X$ $\rightarrow$ $\tt$) at 95$\%$ confidence for $M_X$ between 400 and 1000 $\gevcc$. These limits, as shown in \figref{mxbound}, are used to constrain a model of topcolor assisted technicolor and exclude at 95$\%$ confidence level, the existence of a leptophobic $Z'$~\cite{hillparkeharris} with mass $M_X$$<$560 GeV/$c^2$ for a width $\gammax$ = 0.012$M_{X}$. \\ 
\\

\vspace*{-1.0cm}

%%%%%%%%%%%%%%%%%%%%%%%%%%%%%%%%%%%%%%%%%%%%%%%%%%%%%%%%%%%%%%%%%%%%%%
%  FIGURE 3 Lower bound on M_X {mxbound}.......
%%%%%%%%%%%%%%%%%%%%%%%%%%%%%%%%%%%%%%%%%%%%%%%%%%%%%%%%%%%%%%%%%%%%%%

%-----------------------------------------------------------------------
%  ACKNOWLEDGMENTS
%-----------------------------------------------------------------------
%
We thank the staffs at Fermilab and collaborating institutions, and acknowledge support from the Department of Energy and National Science Foundation (USA), Commissariat  \` a L'Energie Atomique and CNRS/Institut National de Physique Nucl\'eaire et de Physique des Particules (France), Ministry for Science and Technology and Ministry for Atomic Energy (Russia), CAPES and CNPq (Brazil),
Departments of Atomic Energy and Science and Education (India), Colciencias (Colombia), CONACyT (Mexico), Ministry of Education and KOSEF (Korea),
CONICET and UBACyT (Argentina), The Foundation for Fundamental Research on Matter (The Netherlands), PPARC (United Kingdom), Ministry of Education (Czech Republic), A.P.~Sloan Foundation, and the Research Corporation.
%
%-------------------------------------------------------------------------
% References
%-------------------------------------------------------------------------


\vspace*{-0.5cm}
\begin{thebibliography}{0}

\bibitem{topcolor}C. T. Hill and S. Parke, Phys. Rev. {\bf D49}, 
4454 (1994)

\bibitem{hillparkeharris}R. M. Harris, C. T. Hill and S. Parke, Cross section 
for Topcolor $Z'_{t}$ decaying to $\tt$, arXiv:hep-ph/9911288 (1999)

\bibitem{cdf} CDF Collaboration, T. Affolder {\em et al.}, Phys. Rev. Lett. 
{\bf 85 }, 2062 (2000).  

\bibitem{myprl} D$\O$ Collaboration, V.M. Abazov {\em et al.}, {\it Search for narrow width $\tt$ resonances in $\pp$ collisions at $\roots$ = 1.8 TeV}, (to be submitted to Phys. Rev. Lett.). 

\bibitem{bayes}P. C. Bhat, H. B. Prosper, and S. Snyder, Phys. Lett. 
{\bf B407 }, 73 (1997). 

\vspace{-2cm}

\end{thebibliography}
\end{document}